\documentclass[10pt,journal,compsoc]{IEEEtran}

\usepackage[dvipdfm]{graphicx}
\usepackage[T1]{fontenc}
\usepackage[french,british]{babel}
\usepackage{ragged2e}

\usepackage{fancyvrb}
\usepackage{url}
\usepackage{verbatim}
\usepackage{multirow}
\usepackage{pgfplots}
\usepackage{pgfplotstable}
\usepackage{rotating}
\usepackage{amsmath}
\usepackage{booktabs}
\usepackage{threeparttable}
\usepackage{hyperref}
\usepackage{flushend}
\usepackage{blindtext}
\usepackage{setspace}
\usepackage{adjustbox}
\usepackage{algorithm2e}
\usepackage{makecell}
\usepackage{subfigure}
\usepackage{listings}

\ifCLASSOPTIONcompsoc
  \usepackage[nocompress]{cite}
\else
  \usepackage{cite}
\fi

\ifCLASSINFOpdf
\else
\fi

\lstdefinelanguage{xtend}{
  sensitive = true,
  keywords={@Service, @Repository, @Autowired, @Transactional},
  keywords = [2]{public, class, <<, >>,  private, if, else, return, null, true, false},
  keywords = [3]{FOR, ENDFOR, IF, ENDIF},
  keywordstyle=\color{green},
  keywordstyle=[2]\color{blue},
  keywordstyle=[3]\color{black},
  showstringspaces=false,
  breaklines=true,
  frame=top,
  comment=[l]{//},
  morecomment=[s]{/*}{*/},
  commentstyle=\color{purple}\ttfamily,
  stringstyle=\color{red}\ttfamily,
  morestring=[b]',
  morestring=[b]"
  }

\begin{document}
%
\title{Automated Enterprise Applications Generation from
Requirements Model}

\author{
Yilong~Yang
\IEEEcompsocitemizethanks{
\IEEEcompsocthanksitem Yilong Yang is with the School of Software, Beihang University, Beijing, China 100191, and SKLSDE Lab, Beijing, China 100191 (e-mail: yilongyang@buaa.edu.cn).
}

\thanks{Manuscript revised Febrary 19, 2022}}


\IEEEtitleabstractindextext{%
\begin{abstract}
\justifying
Enterprise applications can be automatically generated from a sophisticated OO design model based on model-driven approach. The design model contains information about how to decompose the system into components, how to encapsulate the system operations into classes, and how the objects of classes collaborate to fulfill the functionality of the system operations. However, the efforts to build the design model from a validated requirements model are not proportional to the return. In practice, it is very desirable to have an approach that can automatically generate standardized enterprise applications directly from the validated requirements models. In this paper, we propose an approach named RM2EA, which can reach this goal based on the contract-based requirements model. We demonstrate the proposed approach through 13 case studies. The evaluation result shows that the quality and efficiency of the generated applications are almost equal to the applications implemented by developers: firstly, we demonstrate that a popular type of enterprise applications (i.e., a Jakarta EE application) can be successfully generated by customizing and improving the set of rules; secondly, RM2EA can generate more readable or maintainable code; thirdly, the enterprise applications generated by RM2EA achieve similar performance in test results. Overall, the result is satisfactory, and the implementation of the proposed approach can be further enhanced and applied to software development in the industry.
\end{abstract}

}

\maketitle

\IEEEdisplaynontitleabstractindextext

%
\IEEEpeerreviewmaketitle

\ifCLASSOPTIONcompsoc
\IEEEraisesectionheading{\section{Introduction}\label{sec:introduction}}
\else
\section{Introduction}
\label{sec:introduction}
\fi

\IEEEPARstart{E}{nterprise} Application (EA) \cite{fowler2002z} is often developed as a complete solution for commercial companies, medium and large enterprises, and organizations. The use of enterprise applications facilitates the management of processes within teams and the efficient operations of daily decision-making tasks. However, enterprise applications need to satisfy the business needs of different departments and employees within the company. That makes enterprise applications face the challenges in leveraging speed, security, scalability and reliability. The widely used enterprise development framework such as Jakarta EE (Java EE) and .NET alleviates the challenges, but the developers still need to deal with the complexity of business logic to design and implement reliable enterprise applications. 

Low-code software development \cite{9226356}\cite{khorram:hal-02946812} is a development paradigm, which enables the development of software applications with minimal hand-coding using visual programming with graphical interface and model-driven design. The low-code development tools allow developers to visually deal with complex business logic by facilitating automatic code generation. That is suitable for enterprise application developments. It is very desirable to have a low-code tool to automatically generate a standardized enterprise application (e.g, Jakarta EE applications) directly from requirements model without relying on too much knowledge of architects and developers.

The rapid prototyping method \cite{rapidPrototyping} is to quickly establish a prototype that can reflect users' primary needs. After that, users will make many modifications to the prototype, and developers will quickly modify the prototype system according to the user's opinions. Then give it to the user. Only repeat the above process until the user thinks that the prototype system can meet their requirements. Then, the developer can write the requirements specification according to the prototype system.
Therefore, according to this requirement specification, the developed software should meet the real needs of users. For the developed prototype, because the purpose of the prototype is to know the real needs of users, once the requirements are determined, the prototype will be abandoned. Therefore, the internal structure of the prototype system is not essential. The important thing is to build the prototype quickly and then modify the prototype quickly according to users' opinions.

RM2PT \cite{yang2019automated}\cite{2019RM2PT}\cite{2019RM2PT111} is a low-code approach for rapid prototyping that can automatically transform requirements models into software applications. However, the generated application is a prototype, which is not a fully deployable and reliable enterprise application without support transaction and data persistence to the database.

Nevertheless, RM2PT only generates prototypes, which is far away from an EA. EAs need databases to store persistent data, transactions to keep data safe, need separate front and rear end to respond quickly to changes, and a complete deployment mechanism to make continuous evolution. Moreover, its non-functional requirements are also valued. All of the above mentioned are not well solved by the prototype system. Compared to those work generate code from design, the biggest challenge from prototype to EA comes from requirements, which indicates the final expected effect but contains few details about how to implement the system. Besides, it is also difficult to use design patterns reasonably while realizing the requirements.

In order to further evolve from RM2PT, this paper proposes an OO-based, Non-high concurrent, monolithic EA system generation approach RM2EA, which can automatically generate a standardized enterprise application (i.e., Jakarta EE application) from a contract-based requirements model, and can be deployed to K8S after packaging. The required requirements model can be modeled and validated through RM2PT. The target application can be generated through the proposed transformation rules and algorithm with the generation templates of Jakarta EE. We evaluate our proposed approach RM2EA through 13 case studies by answering four research questions. 

\noindent The contributions of this paper are as follows:
\begin{itemize}
    \item We propose an approach RM2EA that can automatically generate standardized enterprise applications directly from a validated requirements model, and can be deployed in K8S after packaging. This low-code approach contains transformation rules, transformation algorithms and code generation templates of Jakarta EE applications. In the process of generation code, design patterns such as factory pattern are used.
    
    \item We demonstrate that the quality of generated enterprise applications are similar to the applications implemented by developers and the efficiency of generation by comparing with the conventional approach of the manual implementation by the developers.
    
   \item We review requirements modeling and auto-prototyping approach RM2PT and demonstrate it can be used to support enterprise application generation in our proposed low-code approach RM2EA.
\end{itemize}

The remainder of this paper is organized as follows:
Section 2 introduces the related work. Section 3 presents the preliminary of our approach. Section 4 presents the proposed approach RM2EA. Section 5 presents the experiment results with our approach on the 13 case studies. Finally, section 6 concludes this paper and outlines the future work.

\section{RELATED WORK}
Based on the difference between the source and target of generation, the related work contains four parts: 1) generation enterprise applications, 2) generation other types of software system, 3) generation from a design model, and 4) generation from a requirement model.

\begin{table*}[!htb]
  \caption{Representative Work Comparison}
  \label{Compare}
  \centering
  \begin{adjustbox}{max width=2\columnwidth}
  \begin{threeparttable}
  \begin{tabular}{m{4cm}<{\centering} m{3cm}<{\centering} m{3cm}<{\centering} m{2cm}<{\centering} l}
    \toprule
    Team & Input & Output & Domain & Defect \\

    \midrule

    C. McMillan et al\cite{2012Recommending} 
    & feature descriptions, source code
    & recommended source code 
    & C\#, Java
    & \makecell[l]{finding fully applicable source code is often unrealistic,\\ and fully understanding other developers’ code is hard}\\
    
    \midrule
    
    RAPPT\cite{barnett2015bootstrapping}
    & DSL
    & mobile app
    & Android 
    & it only generates the framework of the application developers’ code \\

    
    \midrule
    
    Stefan et al \cite{pleumann2003model}
    & MDR
    & front end code
    & HTML
    & not suitable for developing enterprise applications \\
    
    \midrule
    
    Stenzel et al \cite{2015Formal}
    & QVT
    & Java code
    & Java
    & it doesn't contain complex business logic \\
    
    \midrule
    
    Kundu et al \cite{2013Automatic}
    & UML sequence diagram
    & Java code
    & Java
    & the code is not complete \\


    
    \midrule

    CAYENNE \cite{6945520}
    & requirements documents
    & IEC 61131-3 code
    & Java
    & rules are hard to understand \\
    
    \midrule

    Thramboulidis et al \cite{6059118}
    & P\&IDs, SysML
    & IEC 61131-3 code
    & Java
    & description and implementation have semantic gap \\
    
    \midrule

    Vogel-Heuser et al \cite{1528274}
    & UML model
    & IEC 61131-3 code
    & Java
    & the logic is not complex enough \\
    
    \midrule

    Umple \cite{Forward2012}
    & class model, state machine model
    & application
    & Java, PHP, Ruby
    & it needs developers to complete \\
    
    \midrule

    ActionGUI \cite{ActionGUI2014}
    & design model
    & application
    & Java, Javascript, HTML
    & support for transactions and invariants is not good enough \\
    
    \midrule

    Elkoutbi et al \cite{Elkoutbi2006}
    & UML models
    & GUI application
    & Java
    & models are complex \\
    
    \midrule

    JEM \cite{PASTOR2001507}
    & conceptual model, execution model
    & n-tiered application
    & Java
    & code quality does not meet the needs of enterprise applications \\
    
    \midrule
    
    SCORES \cite{Homrighausen2002}
    & enhancement of the requirements specification
    & application prototype
    & Java
    & \makecell[l]{it is semi-automatic,\\ and sophisticated system operations cannot be generated} \\
    
    \midrule
    
    D. Regep et al \cite{Regep2000}
    & UML models
    & application prototype
    & C++, Ada95
    & it is just code skeleton \\
    
    
    
   \bottomrule
  \end{tabular}

  
  \end{threeparttable}
  \end{adjustbox}
  
\end{table*}

\vspace{.1cm}
\noindent\textbf{Generation Enterprise Applications:} Promoting software reuse by mining feature descriptions and source code from open source repositories is a viable solution \cite{2012Recommending}, But finding fully applicable source code is often unrealistic, and using this approach requires reading and understanding other developers' code, which is a very time-consuming process. RAPPT \cite{barnett2015bootstrapping} is a tool that describes the application in a domain-specific language and eventually implements code generation. However, it only generates the framework of the application. A lot of implementation code is still required to keep the software running. IFML \cite{brambilla2014extending} is a model-driven approach for mobile applications. It enables the automatic generation of mobile applications and has been shown experimentally to save 48\% of the development effort. MDR \cite{pleumann2003model} is an approach to developing Web applications, which supports the development of database-driven applications with limited business logic and close to the underlying entity structure. These studies are useful for improving software development efficiency, but they do not enable the automatic generation of enterprise applications. 

Stenzel \cite{2015Formal} et al. provide a Java code generation framework, but the generated code is basic and does not contain complex business logic.
Kundu \cite{2013Automatic} et al.'s work enables automatic code generation from UML sequence diagrams, but the generated code is not complete.
The study by Sunitha \cite{Sunitha2016Automatic} et al. used activity diagrams and sequence diagrams to model workflows and generate UML models into software applications, but the generated software applications are lacking in front-end user interface-related code.
Rosso \cite{2020Declarative} et al.'s work can be used for web application development, but it uses a custom HTML-like development language to configure components, which requires additional learning costs for users to master.

\vspace{.1cm}
\noindent\textbf{Generation other type of application:} CAYENNE \cite{6945520} can generate standardized programming languages (IEC 61131-3) control logic code from the requirements documents (piping and instrumentation diagrams(P\&IDs)). Thramboulidis and Frey \cite{6059118} sketched a model-driven development process using P\&IDs and SysML to generate IEC 61131-3 code. Vogel-Heuser et al. \cite{1528274} developed a code generator, which can automatically generate IEC 61131-3 code from the UML model. Compared with their work, our method automatically generates standardized enterprise applications directly from requirements models instead of IEC 61131-3 code.

\vspace{.1cm}
\noindent\textbf{Generation from a design model:}
Umple \cite{Forward2012} can generate an application from a class model (conceptual class diagram) and state machine models. However, the state machine only contains abstract states and actions (system operations) descriptions. To generate fully functional applications, programming languages must implement the actions. 
ActionGUI \cite{ActionGUI2014} can generate a multi-tier application from a design model, which includes a data model (specified by ComponentUML  \cite{Basin:2006:MDS:1125808.1125810}), a security model (specified by SecurityUML  \cite{Basin:2006:MDS:1125808.1125810}) and a GUI model (specified by GUI Modeling Language). Our approach can automate EA generation from a requirements model without providing any GUI design. 

The paper \cite{Elkoutbi2006} proposed an intermediate approach to generate GUI applications from UML models. It generates state chart diagrams from a design model specified by a class diagram and collaboration diagrams for each use case. It then generates the UI application from a use case diagram and the intermediate state chart diagram. Compared with their work, we only require a requirements model, which includes a use case diagram, conceptual class diagram (without system operations), system sequence diagrams and the contracts of system operations (only including the interaction between actors and system interface without internal interactions among objects, as required in collaborations diagrams). 

JEM \cite{PASTOR2001507} can generate an n-tiered application from a conceptual model and an execution model. 
SCORES \cite{Homrighausen2002} proposed a semi-automatic approach to generate applications from an enhancement of the requirements specification with a user interface model in FLUID \cite{Kosters1996}. Requirements specification contains a use case diagram, activities diagram to each use case, and a class diagram (including operations). The user interface model includes a specification for view widgets, their navigation, and selection or manipulations (primitive operations) of the domain objects. It does not include the specification or contract for system operations other than simple manipulations of domain objects. Therefore, sophisticated system operations cannot be generated such as \textit{borrowBook}, which includes collaborations of primitive operations such as "find an object", form a link, update the attribute. Moreover, the class diagram in SCORES already contains the system operations in the activity diagram, strictly speaking, that is a design model.

\vspace{.1cm}
\noindent\textbf{Generation from a requirements model:}
The current UML modeling tools, such as Rational Rose, SmartDraw, MagicDraw, Papyrus UML, can only generate skeleton code, where classes only contain attributes and signatures of operations, not their implementations \cite{Regep2000}. The paper \cite{Gemino2004}\cite{Gabrysiak2010} presents a low-code tool, which takes BPMN as input, and generates a mock-up user interface prototype. The generated prototype contains UI pages and navigation to animate workflow execution. Moreover, all interactions of the stakeholders are recorded and incorporated into the prototype models.
The papers \cite{Uchitel04}\cite{Chatley05} present a web animation tool for requirements validation through exploring goals and scenarios, in which it uses linear temporal logic to express goals, and XML to define state transitions and UI components.
TestMEReq \cite{Moketar16} can automatically validate requirements description. It allows multiple stakeholders to validate the same set of requirements collaboratively.

\vspace{.1cm}
\noindent\textbf{Summary:} 
In short, the related previous works 1) require the provision of a design model, which contains a class diagram encapsulating system operations, with the design or implementation of system operations specified in collaboration diagrams or a programming language. Moreover, SCORES, ActionGUI and JEM require a GUI design to generate user interfaces. 2) Only skeleton code or mock-up application can be generated if providing a requirements model. 3) They can not generate standardized enterprise applications such as Jakarta EE or .NET applications.

\section{Preliminary}

In this section, we present some of the techniques related to this paper's approach that will be used to transform requirements models into enterprise-level applications.

\subsection{Requirements Model}
Unified Modeling Language (UML) is the de facto standard for requirement modeling and system design  \cite{larman2012applying}. It contains 1) a use case diagram, which captures domain processes as use cases in terms of interactions between the system and its users, 2) system sequence diagrams, which describe the interactions between actors and system of use case definitions, 3) the contracts of system operations, which specify the conditions that the state of the system is assumed to satisfy before the execution of the system operation, called the {\em pre-condition} and the conditions that the system state is required to satisfy after the execution, called the {\em post-condition} of the system operation, 4) a conceptual class diagram, which contains the conceptual entity classes and their relations of the application domain. Due to the page limitation, the details of requirements model can be found in the RM2PT paper \cite{yang2019automated}. In RM2EA, we mainly use 3) and 4) as input.

\subsection{Automatic Prototyping}
Rapid prototyping is a practical approach to requirements validation to demonstrate concepts, discover requirements errors, and find possible fixing solutions  \cite{Kordon2002}. In practice, it is very desirable to generate prototypes automatically from requirements with a CASE tool. To fill this gap, the papers \cite{yang2019automated}\cite{2019RM2PT}\cite{2019RM2PT111} present an approach and a tool RM2PT to generate prototypes from requirements models automatically. Furthermore, the stakeholders can check whether the requirements reflect their real needs by investigating the executions of use cases in the generated prototypes. 
Besides, the conflict and contradiction of the requirements (especially between the contracts of system operations and invariants) can be automatically detected and further fixed by the features of consistency checking and state observation in the generated prototypes.

\subsection{Development Framework for Enterprise Applications}
Jakarta Enterprise Edition\footnote{\url{https://en.wikipedia.org/wiki/Jakarta_EE}}(Jakarta EE), from Java EE, which is an enterprise-class distributed application development specification jointly developed by many companies under the leadership of Oracle, Inc and Eclipse Foundation. Currently, Jakarta EE \cite{2010Implement} is the dominant solution for enterprise-class distributed application platforms in the market.
With the advancement of Jakarta EE development technology, the Spring\footnote{\url{https://spring.io/projects/spring-framework}} framework is gradually becoming the preferred choice of developers. The Spring Framework provides a comprehensive programming and configuration model for modern Java-based EA. A key element of Spring is infrastructural support at the application level: Spring focuses on the "plumbing" of EA so that teams can focus on application-level business logic without unnecessary ties to specific deployment environments. Spring Boot makes it easy to create stand-alone, production-grade Spring-based Applications. The methodology in this paper is based on the SpringBoot framework to generate an enterprise-level application from the requirements model.

\section{RM2EA Approach}

An enterprise application has to meet the business requirements of different departments and different people within the organization. This leads to the necessity of the EA(Enterprise Application) to contain a large number of business logic implementations that are essential to the EA. In this paper, the implementation code of the business logic is encapsulated in the system operations. The core challenge of EA generation is to generate these system operations. It is generated based on the structural characteristics of the existing software system, thus transforming the requirements model into a coding process, after which the code is complete and ready to run. Among them are the necessity to establish the basic object-oriented enterprise system operations and the transformation rules of OCL expressions to guide the generation process of system operation. This section will first introduce primitive operations for object-oriented enterprise application systems, then introduce the transformation rules used to automate the transformation from OCL expressions into implementation code, and then elaborate on how they work.

\subsection{Overview}
\begin{figure}[!htb]
  \centering
  \includegraphics[scale=0.35]{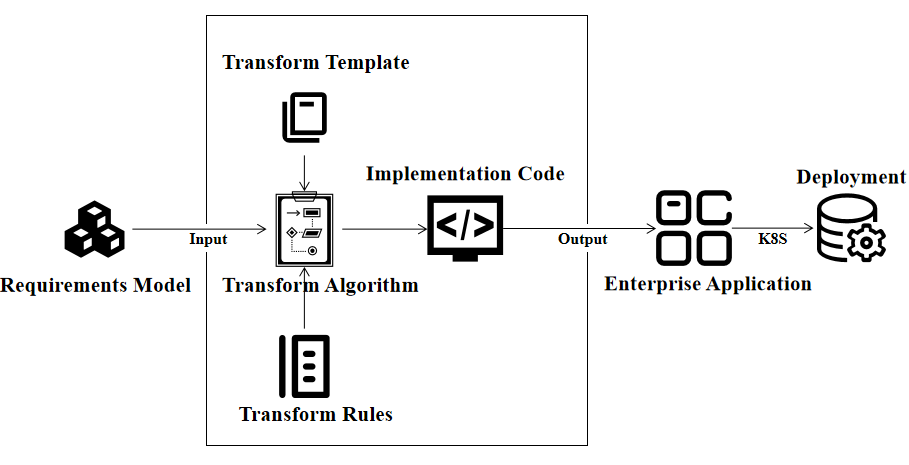}
  \caption{RM2EA Overview}\label{TheoryOfEIS}
\end{figure}

This subsection will give an overview of the principles of EA generation.
A diagram showing the implementation of EA generation in this paper is shown in Figure \ref{TheoryOfEIS}.

The generation method RM2EA takes the requirement model as input and uses a \textbf{transformation algorithm} to process and parse the user requirement information contained in the requirement model. And by using a framework for \textbf{transforming templates} to generate code. For the OCL expressions contained in the system operation contract in the requirements model, developers need to use \textbf{transformation rule} to transform them into the system operation code that drives the running of the EA. The other components of the requirements model are converted into various functional codes to assist in the functioning of the enterprise applications. Finally, the method outputs a complete and executable EA, which can be further deployed in K8S. The subsequent sections will describe in detail the transformation algorithm used to process and parse the requirements model, the transformation template used to generate the code framework, the transformation rule used to generate the system operation code, how they elaborate on the fundamentals of EA generation, and how to further deployed in K8S.

\subsection{OO Primitive Operation}

A set of basic object-oriented operations for object-oriented systems was introduced in RM2PT for system operations. However, since RM2PT works to generate rapid prototypes that can validate requirements, they contain a large amount of implementation code coupled together and does not conform to realistic software development coding standards, nor does it consider the possibility of future application to industry. Therefore, these primitive operations have to be updated to produce software applications that conform to enterprise coding standards. This paper uses the concept of the repository pattern \cite{evans2004domain}, in which the operations on objects in the business logic are transformed into operations on the data warehouse. This allows the core business logic to no longer include the implementation of the data access logic, but rather a dedicated data access layer to perform the data access operations, thus optimizing the business logic layer of the software code and improving the readability and maintainability of the code.
Therefore, the methodology of this paper relies on JPA \cite{H2012Java} query rules based on attribute name, and by analyzing the business logic code of existing enterprise-level systems. We have refined and built seven primitive operations.

The basic operation specification is based on the repository pattern predefined interface, which contains the following main types. The custom operation specification needs to be constructed based on the properties and association relationships of the entity class. In traditional CRUD (create, read, update and delete), \textit{create} can be realized by R4; \textit{read} can be realize by R1, R2, R3 and R8; \textit{update} can be indirectly realize by R5 and R6; \textit{delete} can be indirectly realize by R5 and R7;
\begin{itemize}
\item[(1)] \textbf{save} : 
The save operation returns a boolean value by specifying an object that saves a data object generated by a user operation to the database.
\item[(2)]\textbf{update}:
The update operation returns a boolean value by specifying an object that will be used to update the data after the user has made changes to the data information carried by the object.

\item[(3)]\textbf{delete}:
When specifying an object, the data corresponding to the object in the database is deleted after executing the delete operation, and the return value is a boolean value.

\item[(4)]\textbf{findById}:
Specify the unique identifier of the object, use this operation to find the data in the database based on this unique identifier, the return value is an entity object.

\item[(5)]\textbf{findAll}:
Use this operation to get all the data of a type of collection in the system, and the return value is a collection of objects.

\item[(6)]\textbf{findByAttribute}:
Find the data in the system by using the properties of the object as the data search condition, and the return value is the set of objects.

\item[(7)]\textbf{findByLink}:
Find the data in the system by using the association relationship between objects as the data search condition, the return value is the set of objects.

\end{itemize}

\subsection{Transformation Rule}

RM2EA is improved from RM2PT. Transformation rules defined in RM2PT are used to convert OCL expressions into implementation code, and their main purpose is to generate the business logic layer of the software.
Since those rules defined in RM2PT are only applicable to the generation of prototype systems for monolithic architectures, this paper requires some updates to the transformation rules to generate enterprise-level applications.


Next, we present these updated transformation rules in the form shown below:

\[
\begin{array}{l}
    \textit{Rule}: \frac{\textit{OCL Expression}}{\textit{Primitive Operation in Java Code}}
\end{array}
\]

The transformation rule contains two parts: the top part is an OCL expression, and the bottom part is a primitive operation in java code. In the following transformation rules, \textbf{RepoManage} (Short for RM) is a factory class object, and the \textbf{getRepo()} method of this class can get the repository corresponding to the class name \textit{ClassName}, factory pattern and singleton pattern are used here to realize the data interaction between the class object and the repository.

\[
\begin{array}{l}
    \mathbf{R_1}:

\frac{\textit{obs$\mathord{:}$\textbf{Set}(ClassName)=ClassName.\textbf{allInstances}()}}{\emph{List<ClassName> obs = RM.getRepo(ClassName$\mathord{:}$String).\textbf{findAll}()}} \\  
    
\end{array}
\]

In OCL, The operation \textit{allInstances()} context 
returns all instances of the classifier and the classifiers specializing it.
The rule $R_1$ shows: find all the objects \textit{obs} of the class named \textit{ClassName} in the system. This OCL expression is mapped to the primitive operation \textit{findAll()}, and the found objects are assigned to a list reference \textit{obs} of the class \textit{ClassName}.

\[
\footnotesize
\begin{array}{c}
\mathbf{R_2}: 
\frac{\textit{ob$\mathord{:}$ClassName = ClassName.\textbf{allInstances}()$\rightarrow$\textbf{any}(o $|$ \textbf{conditions}(o))}}{\textit{ClassName ob = RM.getRepo(ClassName$\mathord{:}$String).\textbf{findbyAttribute}(conditions(o)$\mathord{:}$String)}}\\  
    
\end{array}
\]

In OCL, The operation \textit{any()} returns the result of an unspecified value expression in a randomly selection.
The rule $R_2$ are introduced to find an object \textit{ob} from all the instances of the class named \textit{ClassName} with the constraints \textit{condition(o)} by using OCL keyword \textit{any}.
The OCL expressions of $R_2$ are mapped to the primitive operation \textit{findbyAttribute()}, and then assigns the found object to a reference \textit{ob} of class \textit{ClassName}. 

\[
\footnotesize
\begin{array}{c}
    \mathbf{R_3}: 
    
\frac{\textit{ob$\mathord{:}$ClassName = ClassName.\textbf{allInstances}()$\rightarrow$\textbf{any}(o $|$ \textbf{o.id=Id})}}{\textit{ClassName ob = RM.getRepo(ClassName$\mathord{:}$String).\textbf{findbyId}(conditions(Id)$\mathord{:}$Integer)}}\\  

\end{array}
\]

In RM2EA, attributes need to be specifically compared are explicitly placed on the method name to correspond to fields in the database. For example, if we want to fetch a model by the attribute id, the method will be \textbf{findById}. Because this rule is often used, it is extracted separately.

The rule $R_3$ shows: find an object of a class that meets the requirements by the specified ID.
This rule means that in all instances of the class \textit{ClassName} , the object \textit{ob} is found by using the query condition Id using the OCL keyword any.
This OCL expression is mapped to the Primitive operation \textit{findbyId()} and then assign the found object to a reference \textit{ob} of class \textit{ClassName}. 
(in the predefined interface, the findById operation wraps the return value, so it needs to be unwrapped using \textit{GetData()} in the transformation rule).

\[
\begin{array}{c}
    \mathbf{R_4}: 
  
\frac{\textit{\textbf{let} ob$\mathord{:}$ClassName \textbf{in} ob.\textbf{oclIsNew}()}}{\emph{ClassName ob = new ClassName()}}

\end{array}
\]

In OCL, The operation \textbf{oclIsNew} evaluates to true if the object is created during performing the operation.
The rule $R_4$ shows: the object \textit{ob} was created after the execution of system operation.
Mapped code use the basic object-oriented approach of the java language to create the object \textit{ob} of the class named as \textit{ClassName}.
(This OCL expression does not map to the primitive operations mentioned above, but is a new rule built to accommodate changes in software architecture)
\[
\footnotesize
\begin{array}{c}
    \mathbf{R_5}: 
\frac{\textit{ClassName.\textbf{allInstances}()$\rightarrow$\textbf{includes}(ob)}}{\textit{RM.getRepo(ClassName$\mathord{:}$String)}.\emph{\textbf{save}(ob$\mathord{:}$Class)}}

\end{array}
\]
In OCL, \textbf{includes} returns true if object is an element of self, false otherwise.
This OCL expression indicates that the object list of class \textit{ClassName} includes the object \textit{ob} after the execution of the system operation.
Thus, the rule $R_5$ shows create an object \textit{ob} of the class named \textit{ClassName} in the system . This OCL expression is mapped to the primitive operation \textit{save()}.

\[
\begin{array}{c}
    \mathbf{R_6}:

\frac{setAttribute \;of\; ob(ClassName)}
{\textit{RM.getRepo(ClassName$\mathord{:}$String)}.\emph{\textbf{update}(ob$\mathord{:}$Class)}}
    
\end{array}
\]

The attribute setting of OCL is contained in the change of pre-condition and post-condition.
The attributes of the object need to be updated to the system after the assignment. So, the rule $R_6$ shows update an object \textit{ob} of the class named \textit{ClassName} in the system. 
 This OCL expression is mapped to the primitive operation \textit{update()}.
 
\[
\begin{array}{c}
    \mathbf{R_7}: 

\frac{\textit{ClassName.\textbf{allInstances}()$\rightarrow$\textbf{excludes}(ob)}}
{\textit{RM.getRepo(ClassName$\mathord{:}$String)}.\emph{\textbf{delete}(ob$\mathord{:}$Class)}} 
    
\end{array}
\]

The rule $R_7$ shows: delete the object \textit{ob} of the class named \textit{ClassName} in the system object list to make the system state conforming the post-condition. This OCL expression is mapped to the primitive operation \textit{delete()}.

\[
\footnotesize
\begin{array}{c}
    \mathbf{R_8}: 
    
\frac{\textit{obs$\mathord{:}$\textbf{Set}(ClassName)=ClassName.\textbf{allInstances}()$\rightarrow$\textbf{select}(o $|$ \textbf{conditions}(o))}}{\emph{List<ClassName> obs = RM.getRepo(ClassName$\mathord{:}$String).\textbf{findbyAttribute}(conditions(o)$\mathord{:}$String)}}

\end{array}
\]
A \textbf{select} is an operation on a collection, it  specifies a subset of a collection.
The rule $R_8$ shows: find all the objects \textit{obs} of the class named \textit{ClassName} in the system
with the constraints \textit{condition(o)} by using OCL keyword \textit{select}. This OCL expression is mapped to the primitive operation \textit{findObjects()}, and the found objects are assigned to a list reference \textit{obs} of the class \textit{ClassName}.
The OCL expressions of $R_8$ are mapped to the primitive operation \textit{findbyAttribute()}, and then assign the found objects to a reference list \textit{obs} of class \textit{ClassName}. 

\[
\footnotesize
\begin{array}{c}
    \mathbf{R_9}: 

\frac{\textit{obs$\mathord{:}$Set(ClassName) = ob.assoName$\rightarrow$\textbf{select}(o $|$ \textbf{conditions}(o))}}{\emph{List<ClassName> obs = RM.getRepo(ClassName$\mathord{:}$String).\textbf{findbyLink}(ob$\mathord{:}$Class)}}
    
\end{array}
\]

The rules $R_9$ show: find the linked object through \textit{ob.assoName}. 
The OCL expressions of $R_8$ are mapped to the primitive operation \textit{findbyLink()}, and then assign the found objects to a reference list \textit{obs} of class \textit{ClassName}. 

\subsection{Transformation Template}
The requirements model is used as a special notation to describe user requirements and does not define code-level data information. Therefore, to generate the enterprise application implementation code, a transformation template needs to be defined to generate the output style and skeleton of the code. 
In order to generate enterprise-level applications,
we first isolate the code for implementing the Entity Repository with entity manager, system operation, and RESTFul service functional modules by analyzing the component code of the enterprise application (For the entity class section, there is a similar generation process in RM2PT, which is not repeated in this article). Then the default keywords and other reserved symbols are kept, serving as the transformation template's static components. Finally, script processing code is written using Xtend syntax as a dynamic portion of the transformation template and makes it override the custom part of the code, finalize the transformation template for each functional module to build its implementation code.

\subsubsection{System Operation Template}

The template used to generate the implementation of system operation from the contract \textit{c} is shown as follows:
\lstset{
    basicstyle=\footnotesize,
    keywordstyle=\color{purple}\bfseries,
    identifierstyle=\color{brown!80!black},
    commentstyle=\color{gray},
    showstringspaces=false,
    tabsize=2
}


\begin{lstlisting}[language=xtend]
/* Class Skeleton */
@Transactional
@Service
public class <<c.name>> {
    /* Reference object of Repository Manager */
    @Autowired private RepoManage RM;
    /* Parsing system operation contracts */
    <<FOR c : Contract>> 
    /* Generating system operation signatures */
    public <<c.op.returnType>> <<c.op.name>>
        (<<Parameter(c)>>) {
        /* Parsing Contract Definition */
        <<compileDef(resource,c)>> 
        /* Parsing Contract Precondition*/
        <<compilePre(resource,c)>> 
        /* Parsing Contract Postcondition*/
        <<compilePost(resource,c)>> 
    }
    <<ENDFOR>> 
    /* Methods used for data parsing */		
    public static Object GetData(Optional<?> op) {
        if (op.isPresent())
            return op.get();
        else 
            return null;
    }
}
\end{lstlisting}

The template defines a public class with the keyword \textit{public} and the name \textit{c.name} inside of << >> .
Inside the class, the RM object for calling the data repository is defined;
<< FOR>>  and << ENDFOR>>  define the processing scripts for the traversal of the system operation contract;
<< c.op.returnType>> , << c.op.name>> , << Parameter(c)>>  for
system operation return value, signature, and input parameter generation.

\subsubsection{Entity Repository with Entity Manager Template}
The repository function is used to allow for the storage of data and the following template is used to generate the skeleton of this function.

\begin{lstlisting}[language=xtend]
/* Class Skeleton */
@Repository
public interface <<e.name+"Repository">>  extends 
	JpaRepository< <<e.name>>, Integer>,
	JpaSpecificationExecutor< <<e.name>> >{
    
    /* Class Skeleton */
    <<FOR r : e.reference>> 
        <<IF r.isIsmultiple===false>> 
            /* Class Skeleton */
    	    <<IF OneToOne(resource,e.name,r)==false>> 
    		    List< <<e.name>> >findBy<<r.name>>
    		    (<<parm(r)>>);
    	    <<ENDIF>> 
    	    /* Class Skeleton */
    	    <<IF OneToOne(resource,e.name,r)==true>> 
    		    <<e.name>> findBy<<r.name>>
    		    (<<parm(r)>>);
    	    <<ENDIF>> 
        <<ENDIF>> 
    <<ENDFOR>> 
    
    /* Class Skeleton */
    <<compileDaoFind(e,resource)>> 
}

\end{lstlisting}

In the template, the expression defined in << e.name >>  is used to resolve the class name in the concept class diagram. The keyword \textit{public} , \textit{interface}  and the name \textit{e.name+"Repository"} inside of << >>  forms the skeleton of the interface, and inherit the system interface using the extends keyword. The rest of the template will be described in detail in subsequent sections.

\subsubsection{K8S Deployment File Template}
A complete enterprise application may use several layers of architecture, including the front-end application layer, data interface layer, application logic layer, database layer, etc. Furthermore, each layer of architecture may also involve several modules and the corresponding deployment logic process for these modules and layers. So how can the deployment process defined in the development phase be quickly copied to the deployment of other environments, so that the deployment process is automatic and repeatable?

Kubernetes(often abbreviated as K8S) is a container cluster management system. It is an open-source platform, which can realize the functions of automatic deployment, automatic capacity expansion and maintenance of container clusters. K8S can be used to rapidly deploy applications, rapidly expand applications, seamlessly connect new application functions, save resources and optimize the use of hardware resources. 

RM2EA can pack the generated EA and use K8S to deploy with a few steps configured by the K8S deployment file. The configuration file is mainly composed of two parts: docker configuration files and a packaging configuration file. Docker configuration files are used to specify the details for each container, and they are basically fixed. Packaging configuration files integrate the containers as a whole and assign network addresses. They just need to collect the configuration filled in by the user and put them in fixed locations. Take one of the template files as an example, as follows. Here, \textit{<<dbport>>} and \textit{<<webport>>} are database-port and web-port collected from the user.

\begin{lstlisting}[language=xtend]
apiVersion: v1
kind: Service
metadata:
  labels:
    io.kompose.service: mysql
  name: demo
spec:
  ports:
    - name: "3306"
      port: 3306
      nodePort: <<dbport>>
      targetPort: 3306
    - name: "8080"
      port: 8080
      nodePort: <<webport>>
      targetPort: 8080
    - name: "6379"
      port: 6379
      targetPort: 6379
  selector:
    io.kompose.service: mysql
  type: NodePort
\end{lstlisting}

\subsection{Transformation algorithm}

The transformation algorithm can populate the transform template of a functional module by parsing the user's requirements information from the requirements model. Then the complete code is generated, which mainly has two types: the first one is the transformation algorithm of the system operation module, this type of transformation algorithm needs to combine the transformation rule mentioned above, transform the OCL expressions in the operation contract into lines of code according to the restrictions of the transform rules, then encapsulate the lines of code into the implementation class, and finally output to the code file to form a complete system function module. The second type is the transformation algorithm for entity repository with entity manager, RESTFul service, and other functional modules. This type of transformation algorithm is used to improve the functionality of enterprise applications without complex business logic, and only needs to parse the user requirements information from the requirement model to fill in the template, without defining additional rules to assist the code transformation.

\begin{algorithm}
\setstretch{1}
\SetKwData{Left}{left}\SetKwData{This}{this}\SetKwData{Up}{up}
\SetKwFunction{Union}{Union}\SetKwFunction{FindCompress}{FindCompress}
\SetKwInOut{Input}{Input}\SetKwInOut{Output}{Output}

\Input{\textit{ucd} - Use Case Diagram, \\
\textit{ssds} - System Sequence Diagrams, \\
\textit{CS} - Contracts, \\
$t_{so}$ - System Operation Template}
\Output{System Operation Classes}

\Begin{ 
\tcc{Generating use case operation Sets by 
\textit{ucd},
\textit{ssds} and
\textit{contracts}
} 

    Sets$\leftarrow$genera(ucd,ssds,CS)
    
    \For{contract $\in$ CS}{
    	c $\leftarrow$ findContract(SetS)\\

    	 \If{c == contract}{
    	    Generate skeleton by $t_{so}$\\
    	    
    	    OCLexp $\leftarrow$ parse(Contract)
    	    
    	    \For{ sub-exp $\in$ OCLexp}{
    
    	        \If{sub-exp.include(temp)}{
    	            Generate reference object
    	        }
    	    }
    	    
            \For{ sub-exp $\in$ OCLexp}{
                \tcc{Matching transformation rules}
                type $\leftarrow$ Match(sub-exp)\\
                \tcc{Generate code}
                code $\leftarrow$ Generate(type,sub-exp)\\
                \tcc{Encapsulate code in skeleton}
                code $\rightarrow$ skeleton
    	    }    	    
    	 }
    }
  }
  \caption{Generation Algorithm for System Operation}
 \label{atomicAttriandAsso1}
\end{algorithm}

\begin{algorithm}
\setstretch{1}
\SetKwData{Left}{left}\SetKwData{This}{this}\SetKwData{Up}{up}
\SetKwFunction{Union}{Union}\SetKwFunction{FindCompress}{FindCompress}
\SetKwInOut{Input}{Input}\SetKwInOut{Output}{Output}

\Input{\textit{ccds} - Conceptual Class Diagram, \\
\textit{CS} - Contracts, \\
$t_{r}$ - Entity Repository with Entity Manager}
\Output{Entity Repository and Manager}

\Begin{
\For{ccd $\in$ ccds}{
    
    Generate skeleton by $t_{r}$\\
    
    \For{association $\in$ ccd}{
    
        \eIf{Is-Multipe(association)==false}{
            Generate \textit{findOneByLinkObject} code
        }{
            Generate \textit{findManyByLinkObject} code
        }
    }  
    \For{exp $\in$ CS}{
        \eIf{exp IsInclude(any)e}{
            Generate \textit{findOneByLinkObject} code
        }{
            Generate \textit{findOneByLinkObject} code
        }
    }    
    \tcc{Encapsulate code in skeleton}
        code $\rightarrow$ skeleton\;
}}
  \caption{Generation Algorithm for Repository Classes}
 \label{atomicAttriandAsso3}
\end{algorithm}

\subsubsection{System Operation Algorithm}

The system operation generation algorithm is shown in Algorithm \ref{atomicAttriandAsso1}, which uses a system operation contract, a transformation template for the system operation, a use case diagram and a system sequence diagram as input and outputs an implementation class for the system operation. The algorithm first generates a collection of use case operations based on the ucd, ssds, and contracts, which can be used to later encapsulate the system operations into an implementation class to achieve reduced code coupling.
Next, the algorithm performs an iterative operation on the contract, where it obtains a contract c from the set of use cases, compares it with the current contract, and if equal, generates a code framework based on the template $t_{so}$. The algorithm then parses the OCL expression in the contract, iterates over the sub-expressions in the OCL expression, and if there is an attribute shared by the system operation in the sub-expression, generates a reference object for that attribute. Finally, the algorithm performs another iteration of the OCL expression, matches the transformation rule with the type of the sub-expressions, and generates implementation code based on the transformation rule, which is then encapsulated in the code skeleton to enable the generation of the system operation implementation class.

\subsubsection{Entity Repository with Entity Manager Algorithm}

The Repository module generation algorithm is shown in Algorithm \ref{atomicAttriandAsso3}.
the algorithm uses a transformation template, a system operation contract and a conceptual class diagram as input, with the implementation code of the repository module as output. The algorithm starts by iterating over the conceptual class diagram to generate the code framework, and then the algorithm has to execute two types of traversal. The first will iterate over the association of classes, creating a query operation that returns a collection of objects if the association is many-to-one, otherwise creating a query operation that returns a single object. The second way will iterate through the OCL expressions in the system operation contract, search for OCL expressions containing "select" and "any", and determine the type of the expression, if the expression belongs to the "any" operation in the iterative operation, then create a query operation based on the object property name, the return value of which is a single object, otherwise create a query operation based on the object property name, the return value of which is a collection of objects, and finally encapsulate the generated code into the skeleton.

\subsubsection{RESTful Service Algorithm}

The above describes the details of code generation. After fragments are generated, they need to be organized into an EA with RESTful services.
We concatenate the service name and contract name as the API path, and decide what type of request it is, as shown in Algorithm \ref{judgeRequestType}.

\begin{algorithm}
\setstretch{1}
\SetKwData{Left}{left}\SetKwData{This}{this}\SetKwData{Up}{up}
\SetKwFunction{Union}{Union}\SetKwFunction{FindCompress}{FindCompress}
\SetKwInOut{Input}{Input}\SetKwInOut{Output}{Output}

\Input{\textit{pc} - Post Condition}
\Output{Request Type}

\Begin{ 
\tcc{Determine request type by the keyword in \textit{pc}
} 
    \If{pc == null}{
        \Return "GET"
    }
    
    \For{operation $\in$ pc}{
    	putCnt $\leftarrow$ 0 \\
    	 \If{operation contains "excludes"}{
    	   \Return "DELETE"
    	 } 
    	 \If{operation contains "includes}{
    	   \Return "POST"
    	 }
    	 \If{operation contains "self"}{
    	   putCnt ++
    	 }
    }
    
    \If{putCnt != 0}{\Return "PUT"}
    \Return "GET"
  }
  \caption{Generation Algorithm for Request Type}
 \label{judgeRequestType}
\end{algorithm}

\section{Evaluation}
In this section, we will evaluate the proposed approach by answering four research questions.
\subsection{Research Questions}
This paper focuses on answering the main question \textit{"Can enterprise applications be generated directly from requirements models"}? To answer this question, we must first figure out what kind of requirements model is required to generate the EA. Is the proposed approach good enough to generate standardized enterprise applications effectively and correctly? These questions are about soundness, completeness, and efficiency and effectiveness of the proposed approach. The following four research questions are raised based on the main question above:

\vspace{.1cm}
\noindent\textbf{RQ1:} \textit{Are these case studies complex enough to demonstrate the validity of the proposed approach?}

\noindent\textbf{RQ2:} \textit{Can system operations of enterprise applications be generated from a validated requirements model successfully?}

\noindent\textbf{RQ3:} \textit{Are the generated system operations from the validated requirements model correct?}

\noindent\textbf{RQ4:} \textit{Is the quality of the generated code good as the manual implementation by developers?}


\subsection{Case Studies}
In this paper, we will use 13 case studies\footnote{\url{https://github.com/LemonForeast/RM2EA_Examples}}  to assess the effectiveness of the proposed approach. These case studies are systems that are widely used in our daily life.

\subsection{Discussion and Limitation}

\subsubsection{Can enterprise applications be generated directly from requirements models?}

The answer is yes for our proposed approach RM2EA as long as the following conditions are met:

\begin{itemize}
    \item Requirements models should conform to the same UML standards as those given by RM2PT \cite{yang2019automated}, i.e., use case diagrams, conceptual class diagrams, system sequence diagrams;
    \item Contract-based OCL expressions (i.e., pre-and-post conditions, and invariants) must be precisely described;
    \item The established architecture (Springboot with Mysql and Redis) is suitable for actual use.
\end{itemize}

\subsubsection{Limitations}

Currently, RM2EA has some limitations in terms of the types of enterprise applications that can be generated. Due to its requirements models conforming to a restricted subset of UML, it is particularly suitable for generating object-oriented information systems. On the other hand, RM2EA may not be applicable to the generation of systems that have many external interactions with the environments, and batching systems that have heavy internal workloads. It is the same in distributed systems. Moreover, our approach focuses on functional requirements instead of non-functional requirements such as timing requirements, dependability, security, and requirements with spatial constraints. That means our approach does not currently support real-time, embedded, and cyber-physical systems. 

This is not surprising because the inherent complexity of these systems requires more complex requirements engineering models such as the goal-oriented models such as the i-star \cite{Yu1997} for modeling requirements dependencies or KAOS  \cite{Lamsweerde1995} for modeling temporal properties of requirements, and problem-oriented models  \cite{Jackson2001} for contextualizing dependability and security requirements for cyber-physical systems.

In terms of development cost, writing correct OCL contracts and validating them with stakeholders may sometimes exceed the effort saved in coding. In addition, those parts of the requirements that cannot be fully automatically generated still need developers to manually third-party APIs.Some intuitive operations, such as sorting, cannot be well expressed in OCL, so they cannot be generated accurately.


Besides, the project structure is relatively fixed, which means developer may take some time to migrate to another production environment.

\section{Conclusion and Future Work}
In this paper, we propose a low-code approach RM2EA to generate Jakarta EE application from the contract-based requirements model automatically. Requirements modeling is based on the current work of RM2PT. To achieve this goal, we propose Jakarta EE code templates, transformation algorithm and rules that is extended from the 26 transformation rules of RM2PT by adding 9 additional transformation rules for the automatic generation of enterprise-level application systems, so as long as the OCL expressions defined in the system operation contract meet the conditions defined by the transformation rules, they can be transformed into code. Writing a system operation contract containing OCL expressions is challenging in building a requirements model. However, as system operation contracts can describe human requirements more precisely than natural language, this work uses system operation contracts as the basic building block for code generation. 
In the future, we will continue to enhance this low-code approach RM2EA by supporting automating the generation of natural language to system operation contracts, so that the cost and time of requirements modeling, especially for writing OCL contracts can be reduced significantly.

\ifCLASSOPTIONcaptionsoff
  \newpage
\fi

\vfill


\end{document}